\documentclass{elsart3p}
\usepackage{epsfig}
\usepackage{graphicx}
\usepackage{epstopdf}
\usepackage{amsmath}
\usepackage{amssymb}
\usepackage{amsfonts}
\usepackage{mathptmx}
\usepackage{eucal}
\usepackage{bm}
\usepackage{color}

\DeclareMathOperator{\re}{\mathop{\mathrm{Re}}}

\DeclareMathOperator{\arctanh}{arctanh}

\newcommand{\Eq}[1]{Eq.~(\ref{#1})}
\newcommand{\Eqs}[1]{Eqs.~(\ref{#1})}

\begin{document}

\begin{frontmatter}

\title{Detection of small exchange fields in S/F structures}

\author[label1]{A.~S.~Vasenko},
\ead{andrey.vasenko@lpmmc.cnrs.fr}
\author[label2]{S.~Kawabata},
\author[label3]{A.~Ozaeta},
\author[label4,label5]{A.~A.~Golubov},
\author[label6,label7,label8,label5]{V.~S.~Stolyarov},
\author[label3,label9]{F.~S.~Bergeret},
\author[label1,label10]{F.~W.~J.~Hekking}
\address[label1]{LPMMC, Universit\'{e} Joseph Fourier and CNRS, 25 Avenue des Martyrs, BP 166,
38042 Grenoble, France}
\address[label2]{Electronics and Photonics Research Institute (ESPRIT),\\
National Institute of Advanced Industrial Science and Technology (AIST), Tsukuba, Ibaraki, 305-8568, Japan}
\address[label3]{Centro de F\'{\i}sica de Materiales (CFM-MPC), Centro Mixto CSIC-UPV/EHU,
Manuel de Lardizabal 5, E-20018 San Sebasti\'{a}n, Spain}
\address[label4]{Faculty of Science and Technology and MESA$^+$ Institute for Nanotechnology,
University of Twente, 7500 AE Enschede, The Netherlands}
\address[label5]{Moscow Institute of Physics and Technology, 141700 Dolgoprudny, Russia}
\address[label6]{Sorbonne Universit\'{e}s, UPMC Univ Paris 06, UMR 7588,
Institut des Nanosciences de Paris, F-75005, Paris, France}
\address[label7]{CNRS, UMR 7588, Institut des Nanosciences de Paris, F-75005, Paris, France}
\address[label8]{Institute of Solid State Physics RAS, 142432, Chernogolovka, Russia}
\address[label9]{Donostia International Physics Center (DIPC), Manuel de Lardizabal 4, E-20018 San Sebasti\'{a}n, Spain}
\address[label10]{Institut Universitaire de France, 103, bd Saint-Michel 75005 Paris, France}

\begin{abstract}
Ferromagnetic materials with exchange fields $E_{\mathrm{ex}}$
smaller or of the order of the superconducting gap $\Delta$ are
important for applications of corresponding (s-wave) superconductor/
ferromagnet/ superconductor (SFS) junctions. Presently such
materials are not known but there are several proposals how to
create them. Small exchange fields are in principle difficult to
detect. Based on our results we propose reliable detection methods
of such small $E_{\mathrm{ex}}$. For exchange fields smaller than
the superconducting gap the subgap differential conductance of the
normal metal - ferromagnet - insulator - superconductor (NFIS)
junction shows a peak at the voltage bias equal to the exchange
field of the ferromagnetic layer, $eV = E_{\mathrm{ex}}$. Thus
measuring the subgap conductance one can reliably determine small
$E_{\mathrm{ex}} < \Delta$. In the opposite case $E_{\mathrm{ex}} >
\Delta$ one can determine the exchange field in scanning tunneling
microscopy (STM) experiment. The density of states of the FS bilayer
measured at the outer border of the ferromagnet shows a peak at the
energy equal to the exchange field, $E = E_{\mathrm{ex}}$. This peak
can be only visible for small enough exchange fields of the order of
few $\Delta$.
\end{abstract}

\end{frontmatter}

%%%%%%%%%%%%%%%%%%%%%%%%%%%%%%%%%%%%%%%%%%%%%%%%%%%%%%%%%%%%%%%%%%%%%%%%%%%%%
\section{Introduction}

As we know from the quantum theory of magnetism the ferromagnetic
metal can be described by the presence of the so called exchange
field, $E_{\mathrm{ex}}$. This field is responsible to many
interesting phenomena in artificially fabricated superconductor/
ferromagnet (S/F) hybrid structures \cite{RevB, RevG, RevV,
Izyumov}. Let us briefly review the essence of the S/F proximity
effect.

Upon entering of the Cooper pair into the ferromagnetic metal it
becomes an evanescent state and the spin up electron in the pair
lowers its potential energy by $E_{\mathrm{ex}}$, while the spin
down electron raises its potential energy by the same amount. In
order for each electron to conserve its total energy, the spin up
electron must increase its kinetic energy, while the spin down
electron must decrease its kinetic energy, to make up for these
additional potential energies in F. As a consequence, the center of
mass motion is modulated and superconducting correlations in the F
layer have the damped oscillatory behavior \cite{Demler, Eschrig}.
If we neglect the influence of other possible parameters of
ferromagnetic metal (like magnetic scattering rate, etc.) the
characteristic lengths of the decay and the oscillations are equal
to $\xi_f = \sqrt{\mathcal{D}_f/E_{\mathrm{ex}}}$, where
$\mathcal{D}_f$ is the diffusion coefficient in the ferromagnetic
metal \cite{RevB}.

The length $\xi_f$ is also the length of decay and oscillations of
the critical current in Josephson S/F/S junctions \cite{SIFS1,
SIFS2}. Negative sign of the critical current corresponds to the
so-called $\pi$-state \cite{Oboznov, Weides1, Weides2, Kemmler,
Robinson}. S/F/S $\pi$-junctions have been proposed as potential
elements in superconducting classical and quantum logic circuits
\cite{logic1, logic2, Feofanov}. For instance, S/F/S junctions can
be used as complementary elements ($\pi$-shifters) in RSFQ circuits
(see Ref.~\cite{RSFQ} and references therein). S/F/S based devices
were also proposed as elements for superconducting spintronics
\cite{Bakurskiy}. Finally, S/F/S structures have been proposed for
the realization of so called $\varphi$-junctions with a $\varphi$
drop in the ground state, where $0 < \varphi < \pi$
\cite{phi1,phi2}.

Presently known ferromagnetic materials have large exchange fields,
$E_{\mathrm{ex}} \gg \Delta$ and therefore short characteristic
length of oscillations, $\xi_f \ll \xi_s$, where $\xi_s =
\sqrt{\mathcal{D}_s/2\Delta}$ is the superconducting coherence
length and $\mathcal{D}_s$ is the diffusion coefficient in the
superconductor. This requires very high precision in controlling the
F layer thickness in the fabrication process of the Josephson
$\pi$-junctions. In already existing S/F/S structures the roughness
is often larger than the desired precision. The way to solve this
problem is to invent ferromagnetic materials with small exchange
fields.

In this paper we review several proposals for ferromagnetic
materials with exchange fields $E_{\mathrm{ex}}$ smaller or of the
order of the superconducting gap $\Delta$. Then based on our results
we propose reliable detection methods of such small exchange fields
in experiments. Another detection method was recently suggested in
\cite{BG}.

%%%%%%%%%%%%%%%%%%%%%%%%%%%%%%%%%%%%%%%%%%%%%%%%%%%%%%%%%%%%%%%%%%%%%%%%%%%%%

\section{Ways to generate small exchange fields}

The easiest way to create small exchange field is to apply an
external magnetic field $B$ to the normal metal lead, in which case
$E_{\mathrm{ex}} = \mu_B B$, where $\mu_B$ is the Bohr magneton.

It may be also an intrinsic exchange field of weak ferromagnetic
alloys. For example, in Ref.~\cite{small_h} were reported exchange
fields for Pd$_{1-x}$Ni$_x$ with different Ni concentration,
obtained by a fitting procedure (see also \cite{Aarts}). Considering
Nb as a superconductor with $\Delta =1.3$ meV, we can estimate the
exchange field in Pd$_{1-x}$Ni$_x$: for 5.5\% of Ni fitting gives
$E_{\mathrm{ex}} =0.11$ meV, which is 0.1 $\Delta$, for 6\% of Ni it
gives $E_{\mathrm{ex}} =0.45$ meV, which is 0.4 $\Delta$, for 7\% of
Ni it gives $E_{\mathrm{ex}} =2.8$ meV, which is 2.2 $\Delta$, and
for 11.5\% of Ni $E_{\mathrm{ex}} =3.9$ meV, which is 3 $\Delta$.

Another promising alloy with small exchange field,
Pd$_{0.99}$Fe$_{0.01}$, was studied in \cite{PdFe1, PdFe2, PdFe3}.

Finally, a small exchange field can be induced by a ferromagnetic
material into the adjacent normal metal layer. In recent proposal
\cite{Cottet}, a thin normal metal layer was placed on top of the
ferromagnetic insulator. It was shown that the ferromagnetic
insulator may induce effective exchange field in the normal metal
layer \cite{Cottet},
\begin{equation}
E_{\mathrm{ex}}^{\mathrm{eff}} = \hbar \mathcal{D} G_\phi \rho/ d,
\end{equation}
where $\mathcal{D}$ is the diffusion coefficient in the normal
metal, $G_\phi$ is a surface conductancelike coefficient for the
normal metal/ ferromagnetic insulator interface, $\rho$ is the
resistivity of the normal metal, and $d$ is the thickness of the
normal metal layer in the direction, perpendicular to the
ferromagnetic insulator surface. The field
$E_{\mathrm{ex}}^{\mathrm{eff}}$ is expected to be much smaller than
the exchange field inside standard ferromagnets. Interestingly, such
exchange field is possible to tune at the sample fabrication stage
since it is inversely proportional to the normal metal layer
thickness $d$. This gives a flexibility with respect to material
constraints. We also note that as $E_{\mathrm{ex}}^{\mathrm{eff}}
\propto G_\phi$, inducing the tunnel barrier at the normal metal/
ferromagnetic interlayer interface, one can further reduce the value
of the effective exchange field.

Below we suggest direct measurements of such small exchange fields.
The detection methods are different in case of the exchange field
smaller, $E_{\mathrm{ex}} < \Delta$, and larger than the
superconducting gap, $E_{\mathrm{ex}} > \Delta$.

We should mention that we propose methods of small exchange field
detection in the ideal case of ferromagnetic layer with homogeneous
magnetization and absence of magnetic and spin-orbit scattering in
contact with a superconductor. However, in case of realistic
ferromagnets situation can be more complicated. We discuss some
possible limitations of the detection in the end of the two
following sections.

%%%%%%%%%%%%%%%%%%%%%%%%%%%%%%%%%%%%%%%%%%%%%%%%%%%%%%%%%%%%%%%%%%%%%%%%%%%%%

\section{Detection of exchange fields smaller than the superconducting gap}\label{llD}

In this section we consider the following SIFN structure: a
ferromagnetic wire F of a length $d_f$ (smaller than the inelastic
relaxation length \cite{Arutyunov, VH}) is attached at $x = 0$ to a
superconducting (S) and at $x = d_f$ to a normal (N) electrode. The
interface at $x = 0$ is a tunnel barrier while at $x = d_f$ we have
a transparent interface. We will show that the subgap differential
conductance of such a structure has a peak at the bias voltage equal
to the exchange field of the ferromagnetic metal in case when
$E_{\mathrm{ex}} < \Delta$ \cite{Ozaeta1, Ozaeta2}. Thus we propose
to determine small $E_{\mathrm{ex}} < \Delta$ in experiments by
measuring the subgap differential conductance of NFIS junctions at
low temperatures.

In this paper we consider the diffusive limit, i.e. we assume that
the elastic scattering length is much smaller than the decay length
of the superconducting condensate into the F region. Here and below
we consider for simplicity $\mathcal{D}_f = \mathcal{D}_s \equiv
\mathcal{D}$ and $\hbar = k_B = 1$. In order to describe the
transport properties of the system we solve the Usadel equation in
the F layer, that in the so called $\theta$-parametrization reads
\cite{Usadel, Belzig}
\begin{equation}
\frac{\mathcal{D}}{2i} \partial_{xx}^2\theta_{f
\uparrow(\downarrow)}=\left(E \pm E_{\mathrm{ex}}\right)
\sinh\theta_{f \uparrow(\downarrow)}\; .\label{Usadel}
\end{equation}
Here the positive and negative signs correspond to the spin-up
$\uparrow$ and spin-down $\downarrow$ states, respectively. Because
of the high transparency of the  F/N interface the functions
$\theta_{f \uparrow(\downarrow)}=0$ at $x=d_f$. While at the
tunneling interface at $x=0$  we use the Kupriyanov-Lukichev
boundary condition \cite{KL}
\begin{equation}
\partial_x\theta_{f \uparrow(\downarrow)} |_{x=0}
=\frac{R_F}{d_f R_T}\sinh[\theta_{f \uparrow(\downarrow)} |_{x=0} -
\Theta_{s}],\label{BC}
\end{equation}
where $R_F$ and $R_T$ are the normal resistances of the F layer and
SF interface, respectively ($R_T \gg R_F$), and $\Theta_{s}=
\arctanh (\Delta/E)$ is the superconducting bulk value of the
parametrization angle in the S layer, $\theta_s$.  Once the
functions $\theta_{f \uparrow(\downarrow)}$ are obtained one can
compute the current through the junction. In particular we are
interested in the Andreev current, i.e. the current  for voltages
smaller than the superconducting gap due to Andreev processes at the
S/F interface.

Due to the tunneling  barrier at the S/F interface the proximity
effect is weak and hence  we linearize  Eqs.~(\ref{Usadel}-\ref{BC})
with respect to $R_F/R_T \ll 1$.  After a  straightforward
calculation we obtain the Andreev current at zero temperature in
this limit \cite{VZK, VBCH},
\begin{align}
I_A &= \frac{W \Delta^2}{4 e R_T}\sum_{j=\pm}\int_0^{eV}
\frac{dE}{\Delta^2 - E^2}\nonumber
\\
&\times\re\left[ \sqrt{\frac{i \Delta}{E +j E_{\mathrm{ex}}}} \tanh
\left( \sqrt{\frac{E + j E_{\mathrm{ex}}}{i\Delta}}
\frac{d_f}{\xi_s} \right) \right],\label{IA}
\end{align}
where $W=\xi_s R_F / d_f R_T$ is the diffusive tunneling parameter
\cite{KL, Chalmers1, Chalmers2}. In the tunneling limit $W \ll 1$.

We evaluate Eq.(\ref{IA}) in   the  long-junction limit, i.e. when
$d_f\gg\xi_f$, and $E_{\mathrm{ex}} \lesssim eV<\Delta$. We obtain
for the Andreev current
\begin{align}
I_A &= \frac{\Delta \xi_s R_F}{e d_f R_T^2}\sum_{j=\pm} \frac{\arctanh ( c_j^+ )
+\arctan ( c_j^- )}{\sqrt{\Delta+j E_{\mathrm{ex}}}},\label{ia_lj}
\\
c_j^+ &= \sqrt{\frac{eV+j E_{\mathrm{ex}}}{\Delta+j E_{\mathrm{ex}}}}, \quad
c_j^- = \sqrt{\frac{eV-j E_{\mathrm{ex}}}{\Delta+j E_{\mathrm{ex}}}}.\nonumber
\end{align}
\begin{figure}[t]
\epsfxsize=7.5cm\epsffile{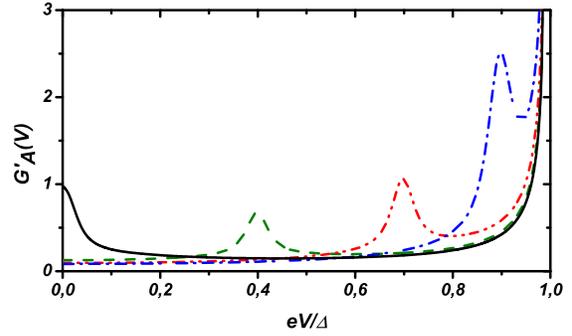} %\vspace{-3mm}
\caption{ (Color online) The bias voltage dependence of differential
conductance at zero temperature for exchange fields,
$E_{\mathrm{ex}}/\Delta=0$ (black solid line),
$E_{\mathrm{ex}}/\Delta=0.4$ (green dashed line),
$E_{\mathrm{ex}}/\Delta=0.7$ (red dash-dot-dotted line), and
$E_{\mathrm{ex}}/\Delta=0.9$ (blue dash-dotted line). Here $G'_A= 2
R_T G_A $, $W=0.014$ and $d_f=10\xi_s$.} \label{GA2D}
%\vspace{-5mm}
\end{figure}
\begin{figure}[t]
\epsfxsize=8cm\epsffile{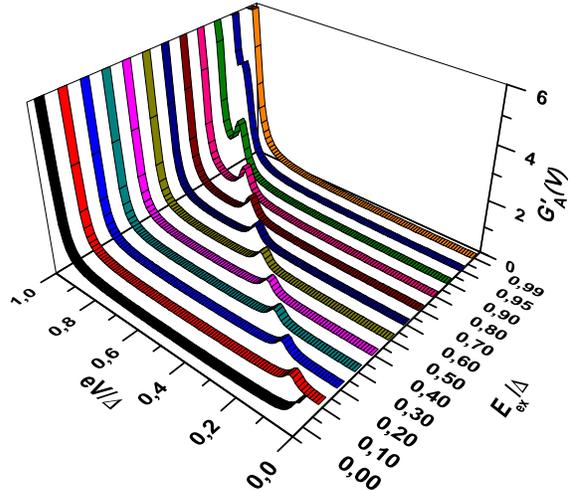} \vspace{-3mm}
\caption{ (Color online) The bias voltage dependence of differential
conductance at zero temperature for exchange fields,
$E_{\mathrm{ex}}/\Delta=0, \; 0.1, \; 0.2, \; 0.3, \; 0.4, \; 0.5,
\; 0.6, \; 0.7, \; 0.8, \;0.9, \; 0.95$ and $0.99$. Here $G'_A= 2
R_T G_A $, $W=0.014$ and $d_f=10\xi_s$.} \label{GA3D} \vspace{5mm}
\end{figure}

In Fig.~\ref{GA2D} we plot the Andreev differential conductance $G_A
= d I_A/ dV$ which is equal to the full differential conductance of
the junction at zero temperature. The conductance shows two well
defined peaks, one at $eV = E_{\mathrm{ex}}$ and the other at $eV =
\Delta$. The detailed physical explanation of the peak at
$E_{\mathrm{ex}}$ is given in \cite{Ozaeta1}. It turns out that it
is the zero bias anomaly (ZBA) peak for the diffusive NIS junction,
shifted in FIS case by $E_{\mathrm{ex}}$.  The ZBA peak in NIS is
shown in Fig.~\ref{GA2D} by black solid line; $E_{\mathrm{ex}}=0$
corresponds to the normal metal case.

In Fig.~\ref{GA3D} we plot the Andreev differential conductance for
many different values of $0 \leq E_{\mathrm{ex}} < 1$ to show the
evolution of the peak with increasing $E_{\mathrm{ex}}$. Detecting
this peak one can carefully measure the value of small exchange
field $E_{\mathrm{ex}} < \Delta$ in the ferromagnetic metal.

We would like to mention that the peak will be visible at $eV =
E_{\mathrm{ex}}$ for a single domain ferromagnet in contact with a
superconductor. In case of a multi-domain ferromagnet the peak of
the differential conductance occurs at $eV$ equals to the
``effective field'', which is the field acting on the Cooper pairs
in the multi-domain ferromagnetic region, averaged over the decay
length of the superconducting condensate into a ferromagnet
\cite{Vasenko_2domains, E_effective}.

%%%%%%%%%%%%%%%%%%%%%%%%%%%%%%%%%%%%%%%%%%%%%%%%%%%%%%%%%%%%%%%%%%%%%%%%%%%%%

\section{Detection of exchange fields larger than the superconducting gap}\label{ggD}

In this section we consider just a simple FS bilayer with a
transparent interface: wire F of a length $d_f$ (smaller than the
inelastic relaxation length \cite{Arutyunov, VH}) is attached at $x
= 0$ to a superconducting electrode by a transparent interface. We
will show that the density of states (DOS) measured at the outer
border of the ferromagnet ($x=d_f$) shows a peak at the energy equal
to the exchange field for $d_f \gg \xi_f$ in case when
$E_{\mathrm{ex}}$ is of the order of few $\Delta$ \cite{SIFS2,
Buzdin_H}. Thus we propose to determine $E_{\mathrm{ex}} > \Delta$
in experiments by measuring the DOS at the outer border of the
ferromagnetic metal in corresponding SF bilayer structure, which can
be done by scanning tunneling microscopy (STM).

\begin{figure}[tb]
\epsfxsize=7.5cm\epsffile{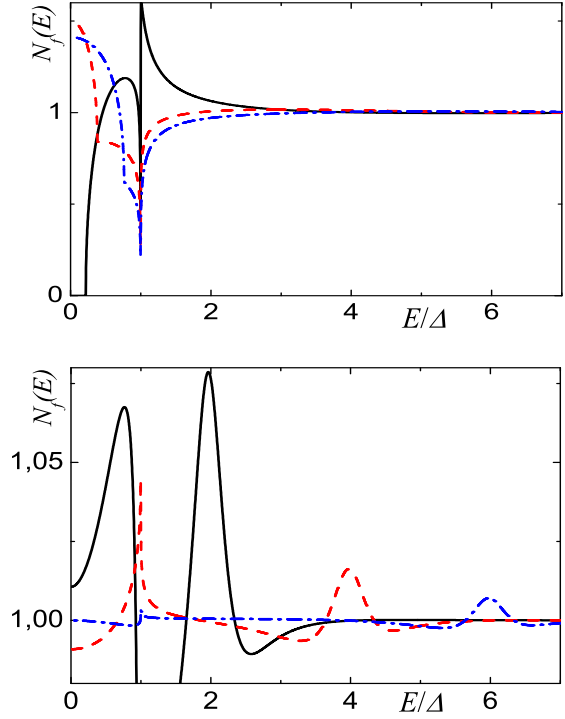} %\vspace{-3mm}
\caption{ (Color online) DOS $N_f(E)$ at the outer border of the F
layer in the FS bilayer calculated numerically for different values
of the exchange field $E_{\mathrm{ex}}$. Parameters of the F/S
interface are $\gamma = \gamma_B = 0.01$, $T = 0.1T_c$. Upper panel:
$d_f/\xi_n = 1$; lower panel: $d_f/\xi_n = 3$. Solid black line
corresponds to $E_{\mathrm{ex}}/\Delta = 2$, dashed red line to
$E_{\mathrm{ex}}/\Delta = 4$, and dash-dotted blue line to
$E_{\mathrm{ex}}/\Delta = 6$.} \label{DOS} \vspace{5mm}
\end{figure}

The DOS $N_f(E)$ normalized to the DOS in the normal state, can be
written as
\begin{equation}
N_f(E) = \left[ N_{f \uparrow}(E) + N_{f \downarrow}(E)\right]/2,
\label{DOS_full}
\end{equation}
where $N_{f \uparrow(\downarrow)}(E)$ are the spin resolved DOS
written in terms of spectral angle $\theta_f$,
\begin{equation}
N_{f \uparrow(\downarrow)}(E) = \re\left[\cosh\theta_{f
\uparrow(\downarrow)}\right].\label{DOS_spin}
\end{equation}
To obtain $N_f$, we use a self-consistent two-step iterative
procedure \cite{SIFS2, triplet}. In the first step we calculate the
pair potential coordinate dependence $\Delta(x)$ using the
self-consistency equation in the S layer in the Matsubara
representation. Then, using the $\Delta(x)$ dependence, we solve the
Usadel equations in the S layer,
\begin{equation}
\frac{\mathcal{D_s}}{2i} \partial_{xx}^2\theta_{s} = E
\sinh\theta_{s} + i\Delta(x) \cosh\theta_s\; ,\label{UsadelS}
\end{equation}
together with the Usadel equations in the F layer [\Eq{Usadel}] and
corresponding boundary conditions, repeating the iterations until
convergency is reached \cite{SIFS2}. At the outer border of the
ferromagnet ($x=d_f$) we have $\partial_x \theta_{f
\uparrow(\downarrow)} = 0$. At $x=0$ we use Kupriyanov-Lukichev
boundary conditions which in case of the transparent interface is
convenient to write as
\begin{subequations}\label{KL1}
\begin{align}
\gamma \partial_x \theta_f |_{x=0} &= \partial_x \theta_s |_{x=0} ,
\\
\xi_n \gamma_B \partial_x\theta_{f} |_{x=0} &=\sinh(\theta_{f} -
\theta_{s})|_{x=0}.
\end{align}
\end{subequations}
Here $\gamma = \sigma_f/ \sigma_s$, $\sigma_{f(s)}$ are the
conductivities of the F (S) layers correspondingly, $\xi_n =
\sqrt{\mathcal{D}/ 2 \pi T_c}$, $T_c$ is the critical temperature of
the superconductor, and $\gamma_B = d_f R_T/\xi_n R_F = \xi_s/ \xi_n
W$. The parameter $\gamma$ determines the strength of suppression of
superconductivity in the S layer near the interface (inverse
proximity effect). No suppression occurs for $\gamma = 0$, while
strong suppression takes place for $\gamma \gg 1$. In our numerical
calculations we assume small $\gamma \ll 1$. Since we consider the
transparent interface $R_F \gg R_T$ and contrary to the previous
section $W \gg 1$, therefore $\gamma_B \ll 1$. Notice that in the
\Eqs{UsadelS}-\eqref{KL1} we have omitted the subscripts
$\uparrow(\downarrow)$ because equations for both spin directions
are identical in the superconductor.

In Fig.~\ref{DOS} we plot the DOS $N_f(E)$ at the outer border of
the F layer in the FS bilayer calculated numerically for different
values of the exchange field $E_{\mathrm{ex}}$ and for different F
layer thicknesses $d_f$. At large enough $d_f$ ($d_f/\xi_n = 3$) we
see the peak at $E = E_{\mathrm{ex}}$ [see Fig.~\ref{DOS}, lower
panel]. For small $d_f$ ($d_f/\xi_n = 1$) the peak is not visible
and DOS tends monotonously to unity for $E
> \Delta$ [see Fig.~\ref{DOS}, upper panel]. The amplitude of the
peak is decreasing with increasing $E_{\mathrm{ex}}$: peak is only
visible for $E_{\mathrm{ex}}$ of the order of few $\Delta$ (see
\cite{SIFS2} for details). We also need to mention that in case of
$E_{\mathrm{ex}} < \Delta$ there is no peak in the DOS.

To better illustrate the conditions when the peak at $E =
E_{\mathrm{ex}}$ is visible in experiments we consider an analytical
limiting case. If the F layer is thick enough ($d_f \gg \xi_{f}$)
and $\gamma = 0$ in \Eq{KL1}, the DOS at the outer border of the
ferromagnet can be written as \cite{SIFS1, SIFS2, Cretinon}
\begin{equation}  \label{DOS_bound}
N_{f \uparrow(\downarrow)}(E) = \re[ \cos\theta_{b
\uparrow(\downarrow)} ] \approx 1 - \frac{1}{2}\re \theta_{b
\uparrow(\downarrow)}^2.
\end{equation}
Here $\theta_{b \uparrow(\downarrow)}$ is the boundary value of
$\theta_f$ at $x=d_f$, given by
\begin{equation}  \label{theta_bound}
\theta_{b \uparrow(\downarrow)}= \frac{8 F(E)}{\sqrt{F^2(E) + 1} +
1} \exp\left( -p_{\uparrow(\downarrow)} \frac{d_f}{\xi_f}\right),
\end{equation}
where we use the following notations,
\begin{subequations}
\label{qef}
\begin{align}
p_{\uparrow(\downarrow)} &= \sqrt{2/E_{\mathrm{ex}}}\sqrt{-iE_R \pm iE_{\mathrm{ex}}}, \label{p}\\
F(E) &= \frac{\Delta}{-iE_R + \sqrt{\Delta^2 - E_R^2}}, \quad E_R =
E + i0.
\end{align}
\end{subequations}

From \Eqs{DOS_bound}-\eqref{theta_bound} we obtain for the full DOS
the following expression in the limit  $d_f \gg \xi_{f}$ and for $E
\geq \Delta$,
\begin{align}
N_f(E) = 1 + \sum_{j=\pm} \frac{16 \Delta^2 \cos(b_j) \exp(-b_j)}
{(E+\epsilon)(\sqrt{E+\epsilon} + \sqrt{2\epsilon})^2},\label{Exp}
\end{align}
\begin{equation}
b_j = \frac{2d_f}{\xi _{f}}\sqrt{\frac{|E +j
E_{\mathrm{ex}}|}{E_{\mathrm{ex}}}}, \quad \epsilon = \sqrt{E^2 -
\Delta^2}.\nonumber
\end{equation}
We can clearly see the exponential asymptotic of the peak at $E =
E_{\mathrm{ex}}$ from \Eq{Exp}. We should keep in mind that \Eq{Exp}
is valid for large $d_f/\xi_f$, but nevertheless we may
qualitatively understand why we do not see the peak at $E =
E_{\mathrm{ex}}$ for small ratio of $d_f/\xi_f$: if this factor is
small the variation of the exponent $\exp\{-2(d_f/\xi_{f})\sqrt{|E -
E_{\mathrm{ex}}|/E_{\mathrm{ex}}}\}$ near the point $E =
E_{\mathrm{ex}}$ is also small. The peak is observable only for
$E_{\mathrm{ex}}$ of the order of a few $\Delta$. For larger
exchange fields the peak is difficult to observe, since the energy
dependent prefactor of the exponent in \Eq{Exp} decays as $E^{-2}$
for $E \gg \Delta$.

Detecting this peak one can carefully measure the value of small
exchange filed $E_{\mathrm{ex}} > \Delta$ in the ferromagnetic
metal.

We mention that in the presence of magnetic scattering the DOS peak
at $E = E_{\mathrm{ex}}$ do not change the position but gets smeared
at large enough scattering rate \cite{SIFS2}. We did not consider
the effect of domain structure of the F layer on this peak, but we
can expect similar behavior as discussed in previous section, i.e.
the position of the DOS peak will move to the value of the
``effective filed'' \cite{E_effective}.

\section{Summary}

We propose reliable methods to measure small exchange fields in weak
ferromagnet/ superconductor structures. For $E_{\mathrm{ex}} <
\Delta$ the subgap differential conductance of the normal metal -
ferromagnet - insulator - superconductor (NFIS) junction shows a
peak at the voltage bias equal to the exchange field of the
ferromagnetic layer, $eV = E_{\mathrm{ex}}$. Thus measuring the
subgap conductance one can reliably determine small $E_{\mathrm{ex}}
< \Delta$. In the opposite case $E_{\mathrm{ex}} > \Delta$ one can
determine the exchange field in scanning tunneling microscopy
experiment. The density of states of the FS bilayer measured at the
outer border of the ferromagnet shows a peak at the energy equal to
the exchange field, $E = E_{\mathrm{ex}}$.

Next we are planning to search for small exchange fields
$E_{\mathrm{ex}} > \Delta$ in the experiments, using the ultrahigh
vacuum Scanning Tunneling Microscopy (STM) and Spectroscopy (STS)
technique, recently developed by one of the authors (Stolyarov et
al.) \cite{Stolyarov}.

We also hope that our results will trigger further experimental
activity in finding ferromagnetic materials with small exchange
fields. Good candidates for such materials can be diluted
ferromagnetic alloys (like PdNi, PdFe, CuNi, etc.) and normal metals
in proximity with the ferromagnetic insulators (FI). In the latter
case the ferromagnetic insulator may induce the exchange field in
the thin normal metal layer, placed on top of the FI material.

This work was supported by European Union Seventh Framework
Programme (FP7/2007- 2013) under grant agreement ``INFERNOS'' No.
308850, by Ministry of Education and Science of the Russian
Federation, grants No. 14Y.26.31.0007, No. 2014-14-588-0010-061,
RFBR No. mol\_a 14-02-31798, and by French National Agency for
Research ANR (ELECTROVORTEX). A.S.V. acknowledge the hospitality of
Superconducting electronics group, AIST, during his stay in Japan.

%%%%%%%%%%%%%%%%%%%%%%%%%%%%%%%%%%%%%%%%%%%%%%%%%%%%%%%%%%%%%%%%%%%%%%%%%%%%%

\end{document}